\documentclass[prb,twocolumn,showpacs,superscriptaddress]{revtex4}

\usepackage{epsf}
\usepackage{epsfig}
\usepackage{graphicx}
\usepackage{dcolumn}
\usepackage{braket}
\usepackage{bm}
\usepackage{amsfonts}
\usepackage{amsmath}
\usepackage{amssymb}
\usepackage{color,soul}

\input epsf

\begin{document}

\title{Radiation effects on the electronic structure of bilayer graphene}

\author{Eric Su\'{a}rez Morell}
\affiliation{Departamento de F\'{\i}sica, Universidad T\'{e}cnica Federico Santa Mar\'{\i}a, Casilla 110-V, Valpara\'{\i}so, Chile}

\author{Luis E. F. Foa Torres}
\affiliation{Instituto de F\'{\i}sica Enrique Gaviola (IFEG-CONICET) and
FaMAF, Universidad Nacional de C\'{o}rdoba, Ciudad Universitaria, 5000
C\'{o}rdoba, Argentina.}

\date{\today}

%
\begin{abstract}
We report on the effects of laser illumination on the electronic properties of bilayer graphene. By using Floquet theory combined with Green's functions we unveil the appeareance of laser-induced gaps not only at integer multiples of $\hbar \Omega /2$ but also at the Dirac point with features which are shown to depend strongly on the laser polarization. Trigonal warping corrections are shown to lead to important corrections for radiation in the THz range, reducing the size of the dynamical gaps. Furthermore, our analysis of the topological properties at low energies reveals that when irradiated with linearly polarized light, ideal bilayer graphene behaves as a trivial insulator, whereas circular polarization leads to a non-trivial insulator per valley. 

\end{abstract}

\maketitle

\section{Introduction}

Among the many promises sparked by graphene research during the last few years \cite{Geim2009,CastroNeto2009}, graphene optoelectronics is perhaps one of the brightest \cite{Bonaccorso2010,Xia2009,Karch2011,Konstantatos2012}. From improved power conversion of energy harvesting devices \cite{Gabor2011} to novel plasmonics properties \cite{Koppens2011,Chen2012}, graphene and related materials offer an outstanding playground for te study of light-matter interaction with many potential applications \cite{Bonaccorso2010,Ren2009,Xia2009,McIver2012}.

Recent studies pointed out the intriguing possibility of inducing bandgaps in monolayer graphene by illumination with a laser field \cite{Syzranov2008,Oka2009,Kibis2010}. The peculiar electronic structure of graphene and its low dimensionality are crucial for the occurence of this effect. Further studies have predicted observable changes in the conductance \cite{Calvo2011,Gu2011} and optical properties \cite{Zhou2011} , with a strong dependence on laser polarization \cite{Calvo2011,Savelev2011}, setting off many other interesting studies \cite{Dora2012,Iurov2012,Liu2012,Busl2012,SanJose2012}. Moreover, the possibility of controlling topological insulators with photocurrents \cite{McIver2012}, as well as the emergence of non trivial laser-induced topological properties and edge states \cite{Lindner2011,Kitagawa2011,Gu2011,Dora2012}, the so-called Floquet topological insulators \cite{Lindner2011,Kitagawa2011}, add more relevance to this area.

Graphene's thicker cousin, bilayer graphene (BLG), has also shown an enormous potential \cite{Geim2009,CastroNeto2009}, allowing for a tunable bandgap \cite{Zhang2010} as required for the operation of active devices. Notwithstanding, the studies mentioned in the last paragraph were all centered in monolayer graphene. Only in Ref. \cite{Abergel2009}, the authors proposed irradiated bilayer as a vehicle for inducing a valley polarized current. Here, we focus on the electronic and topological properties of bilayer graphene illuminated by a laser with frequency either in the THz or in the mid-infrared range. In the THz range, trigonal warping (TW) corrections are shown to induce strong modifications in the theoretical predictions leading, besides qualitative changes in the spectra, to quantitative differences in the laser-induced gaps up to a factor of two. 

Moreover, we show that a laser field may also lead to polarization-tunable topological properties in BLG ranging from a trivial insulator to one with properties akin those of a topological insulator. Specifically, we show that the low energy properties of BLG illuminated by circularly polarized light can be described by a simple effective Hamiltonian similar to the one of BLG with a bias. Our theoretical analysis shows that although the system behaves as a trivial insulator in the presence of linearly polarized light, switching the polarization to circular transforms it into a non-trivial insulator per valley.

\begin{figure}[tbp]
\includegraphics[width=8.2cm]{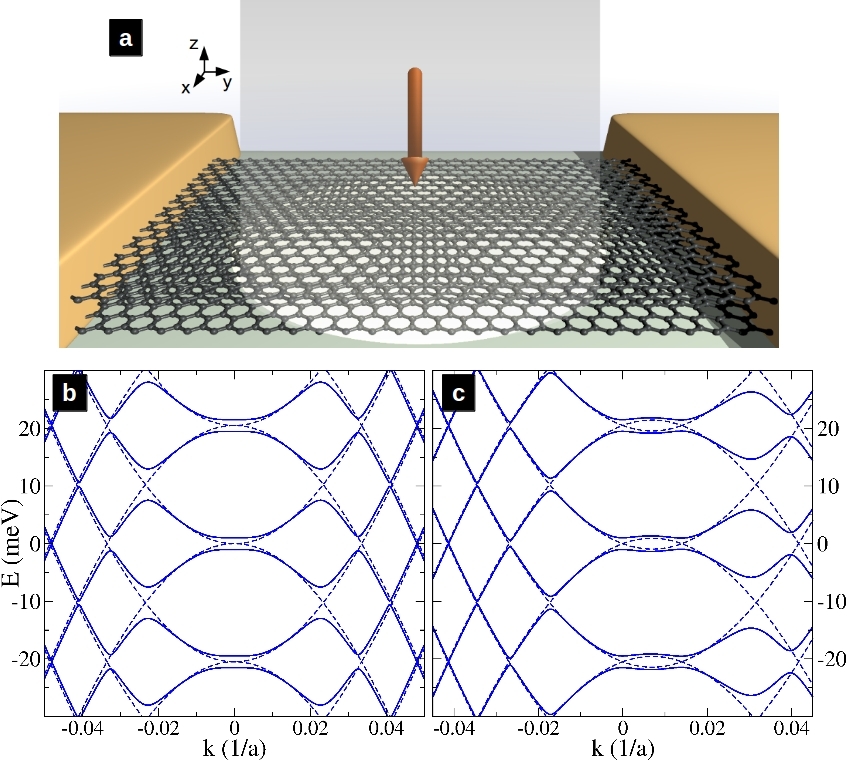}
\caption{(Color online) (a) Scheme of the considered setup where a laser field is applied perpendicular to a graphene bilayer. Panels (b) and (c) show the quasienergy Floquet band structure for bulk bilayer without and with trigonal warping respectively. These plots are along $k_{x}$ direction ($k_{y}=0$), solid lines are for  circularly polarized light with a frequency of $5$ THz and an intensity of $0.5 \, mW/\mu m^2$. The unirradiated spectrum is shown with dashed lines.}
\label{fig-1}
\end{figure}
  
\section{Results and discussion}
\subsection{Floquet theory applied to irradiated bilayer graphene.}

The unit cell of bilayer graphene with Bernal stacking has two inequivalent sites labeled as $A1$, $B1$ on the top layer and $A2,B2$ on the bottom layer, they are arranged in such a way that atom $B1$ lies on top of atom $A2$. Using the wave-functions $\Psi=(\psi_{A1}, \psi_{B2}, \psi_{A2}, \psi_{B1})^T$  for the $K$ valley and $\Psi=(\psi_{B2}, \psi_{A1}, \psi_{B1}, \psi_{A2})^T$ for the $K'$ valley, an effective Hamiltonian for the low energy properties is given by \cite{McCann2006}:

\begin{equation}
H_{0}(\vec{k})=\xi \left( \begin{array}{cccc}   
0 & v_{3} \pi & 0 & v \pi^{\dagger}\\
v_{3} \pi^{\dagger} & 0 & v \pi & 0\\
0 & v \pi^{\dagger} & 0 & \xi \gamma_{1}\\
v \pi & 0 & \xi \gamma_{1} & 0
\end{array} \right),
\label{eq_Hamiltonian} 
\end{equation}
where $\xi=1 (-1)$ for valley $K(K')$, $\pi=p_{x}+ i \ p_{y}$, $v=(\sqrt{3}/2) a \gamma_{0}/\hbar$, $v_{3}=(\sqrt{3}/2) a \gamma_{3}/\hbar$, $a=0.246$ nm, graphene lattice constant, $\gamma_{0}=3.16$ eV, $\gamma_{1}=0.39$ eV and $\gamma_{3}=0.315$ eV. The hopping parameter $\gamma_{3}$ is responsible for the trigonal warping effects. 


We apply linearly/circularly polarized light perpendicular to the graphene bilayer as shown schematically in Fig. \ref{fig-1}a. The time-dependent field is introduced through the substitution $\vec{k}\rightarrow \vec{k}+e \vec{A}/\hbar $, where the vector potential is $\vec{A}(t)=A (cos(\Omega t),cos(\Omega t+\phi))$, where $\phi=0 \, (\pi/2)$ for linear(circular) polarization. The Floquet theorem \cite{Platero2004,Kohler2005,FoaTorres2005} provides an elegant route to handle this time-periodic Hamiltonian ($H(t+T)=H(t)=H_0(\vec{k}+e \vec{A}(t)/\hbar)$, where $T=2 \pi/\Omega$), it states that the solutions to the time-dependent Schr\"{o}dinger equation can be written as $\Psi_{\alpha}(\vec{r},t)=e^{-i \varepsilon_{\alpha} t/\hbar} \phi_{\alpha}(\vec{r},t)$, where $\phi_{\alpha}(t)=\phi_{\alpha}(t+T)$ is time-periodic, the Floquet states can be further expanded into a Fourier series $\phi_{\alpha}(t)=\sum e^{i n \Omega t} \phi_{\alpha}^{(n)}$ and a substitution in the Schr\"{o}dinger equation gives:

\begin{equation}
\sum_{m}(H^{(n,m)}-n \hbar\Omega \delta_{n,m}) \ket{\phi_{\alpha}^{(n)}} = \epsilon_{\alpha} \ket{\phi^{(n)}},
\label{eq_static_Hamiltonian} 
\end{equation}
where  $H^{(n,m)}=\frac{1}{T} \int_{0}^{T} dt H(t) e^{i (n-m) \Omega t}$ and $\epsilon_{\alpha}$ is the so-called quasi-energy. Simple inspection shows that this is an eigenvalue equation analog to the one for time-independent systems. There are however two main differences: the role of the Hamiltonian is played by the so-called Floquet Hamiltonian $H_F=H-i\hbar \, d/dt$; and the states belong to an extended Hilbert space which is the direct product between the usual Hilbert space and the space of time-periodic functions with period $T$. It is straightforward to see that \textit{H}$_{F}^{(n,m)}= H^{(n,m)}-n\hbar \Omega \delta_{n,m}$. This method has been applied to a variety of systems and in particular to ac fields such as alternating gate voltages in graphene \cite{FoaTorres2011,SanJose2011} beyond the adiabatic limit.

The time-averaged density of states (DOS) gives valuable information on the Floquet spectra in a compact form and can be calculated as in \cite{Oka2009,Calvo2011}. To such end we compute the Floquet-Green function, defined as $\mathbf{G}_{F}=\left(\mathbf{\varepsilon1}-\mathbf{H}_{F}\right)^{-1}$, from which the time-averaged DOS is obtained as $ DOS(\varepsilon)=-\frac{1}{\pi}Im\left\{ Tr\left(\mathbf{G}_{F}(\varepsilon)\right)_{0,0}\right\}$, where $(\mathbf{G}_{F})_{0,0}$ stands for the sub-block of the Floquet-Green function corresponding to vanishing Fourier index.

\subsection{Laser-induced modifications of the Floquet spectra} 

In the following we will analyze the behavior of the quasi-energy spectra and the DOS for various laser intensities, frequencies and polarization \cite{note}.

\begin{figure}[tbp]
\includegraphics[width=8.2cm]{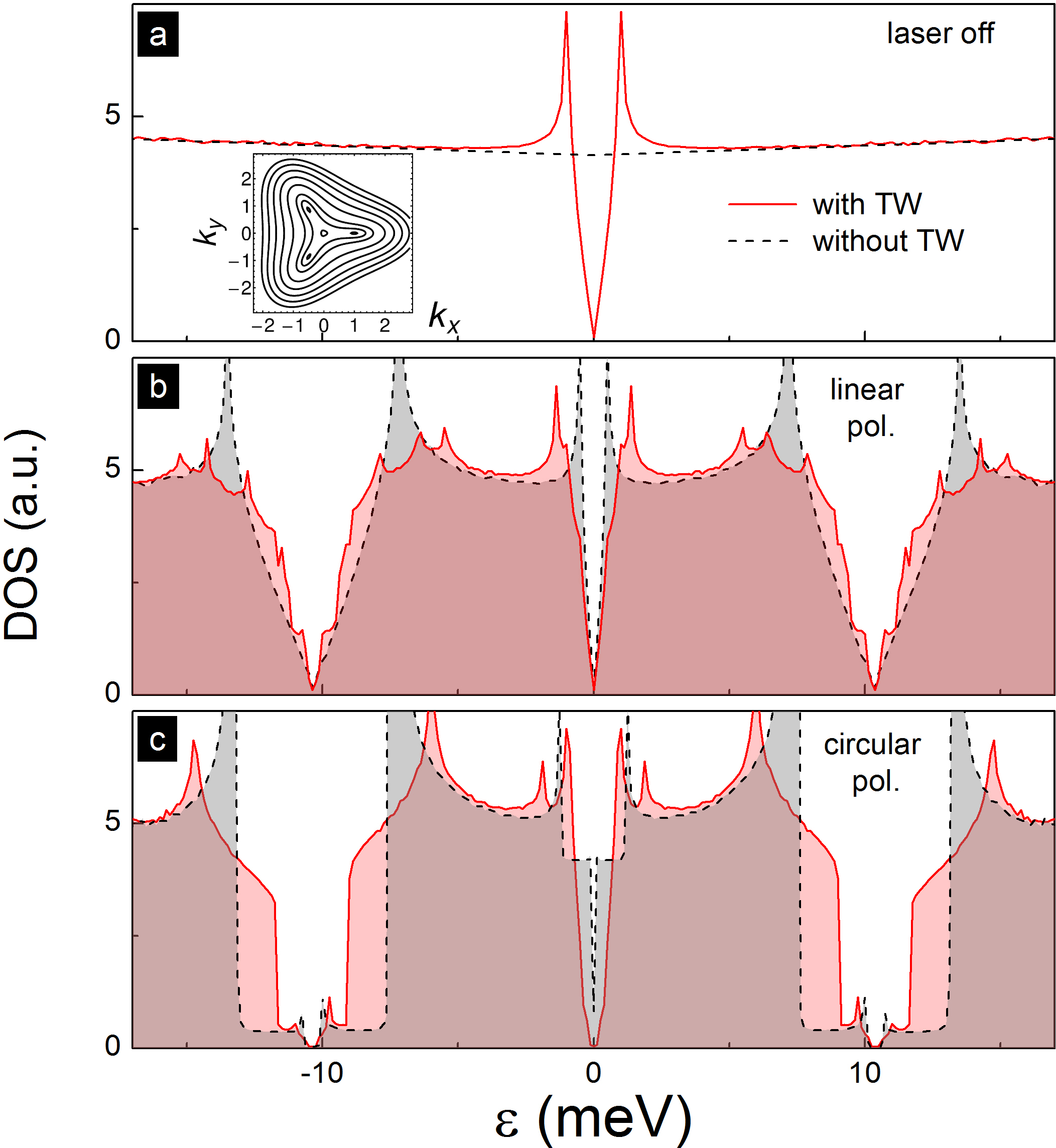}
\caption{(Color online) a) DOS for bilayer graphene with the laser turned off. The continous line corresponds to calculations performed including trigonal warping (TW) corrections and the dashed line without them. Panel a)-inset depicts the iso-energy lines for the dispersion of bilayer graphene in the absence of radiation, the strong TW distortion is evident. Panels b) and c) show the DOS as defined in the text for bilayer graphene in the presence of linearly (b) and circularly (c) polarized light (5 THz) with an intensity of $0.5$ mW/$\mu m^2$.}
\label{fig-2}
\end{figure}

While in monolayer graphene trigonal warping (TW) introduces small corrections which become noticeable only at high energies ($\sim $500 meV), in the case of bilayer graphene these corrections are stronger at low energies where they lead to a splitting of the Dirac point into a structure with four pockets \cite{McCann2006} as shown in the inset of Fig.\ref{fig-2}a. Here we show that these effects, that were neglected in previous studies of irradiated bilayer, are indeed very important for radiation in the THz range. 

Figure \ref{fig-1} shows the quasi-energy dispersion along a particular $k$ direction without (b) and with (c) the TW correction in the presence of the electromagnetic field. The dashed lines in each figure shows the unirradiated case. The field is expected to have the stronger effect at the crossing points which, due to the electron-hole symmetry, are located at integer multiples of $\hbar\Omega/2$ above and below the Dirac point, as can be seen in Fig. \ref{fig-1}b and \ref{fig-1}c. The time-dependent perturbation introduces a non-vanishing matrix element between the states at those crossings, thereby lifting the degeneracies and opening the so called dynamical gaps \cite{Syzranov2008,Oka2009}. The gap at the charge neutrality point is a higher-order effect and will be analyzed in more details later.

Figure \ref{fig-2} shows the DOS for bilayer graphene in the presence of either linearly (b) or circularly (c) polarized light (5THz) with (solid-line) and without (dashed-line) TW. The DOS in the absence of radiation is shown in (a) for reference. Although from the discussion before one may expect the main corrections to arise only close to the Dirac point, Figs. \ref{fig-2}b and \ref{fig-2}c show that they emerge even at the dynamical gaps for radiation in the THz range. 

For linearly polarized light, the DOS in the vicinity of $\hbar\Omega/2$ exhibits a depletion area with a linear dispersion and a single point of vanishing DOS. This is similar to the case of monolayer graphene found in Ref. \cite{Calvo2011} and is due to the fact that the gap depends on relative angle between $k$ and the polarization vector, no gap emerges when they are parallel. One can also notice that the roughly linear dispersion around the dynamical gaps acquire a structure with three narrow features on each side when TW corrections are included. This is a consequence of the deformation of the iso-energy lines in the $k_x$-$k_y$ plane due to the TW corrections (see inset of Fig. \ref{fig-2}a).

For circular polarization, two striking observations not reported before should be emphasized: a) There is a gap opening at zero energy (which also occurs in the absence of TW but is much smaller and cannot be distinguished in the figure, see inset of Fig. 1-b); and b) the dynamical gap (which turns out to be linear in the field intensity as for monolayer graphene) is overestimated by almost a factor two when the TW corrections are not taken into account. A key ingredient behind these differences is again the breaking of the rotational symmetry in the $k_x$-$k_y$ plane even for low energies. Though the gap at the charge neutrality point would require stringent conditions (being of about 0.3 meV for a laser intensity of 0.5mW/micronsquare), the physics described here may prompt additional research and experiments that may allow directly or indirectly to unveil it. In contrast, the effects described at the dynamical gaps ($\pm\hbar\Omega/2$) are much stronger and should be observable in low temperature experiments. Indeed the dynamical gaps are of the order of 5K for 10Thz radiation at $0.5 mW/\mu^2$ and reach larger values (up to 30meV, or 350K) for 30THz radiation for a power of a few $mW/\mu^2$.

\begin{figure}[tbp]
\includegraphics[width=8.2cm]{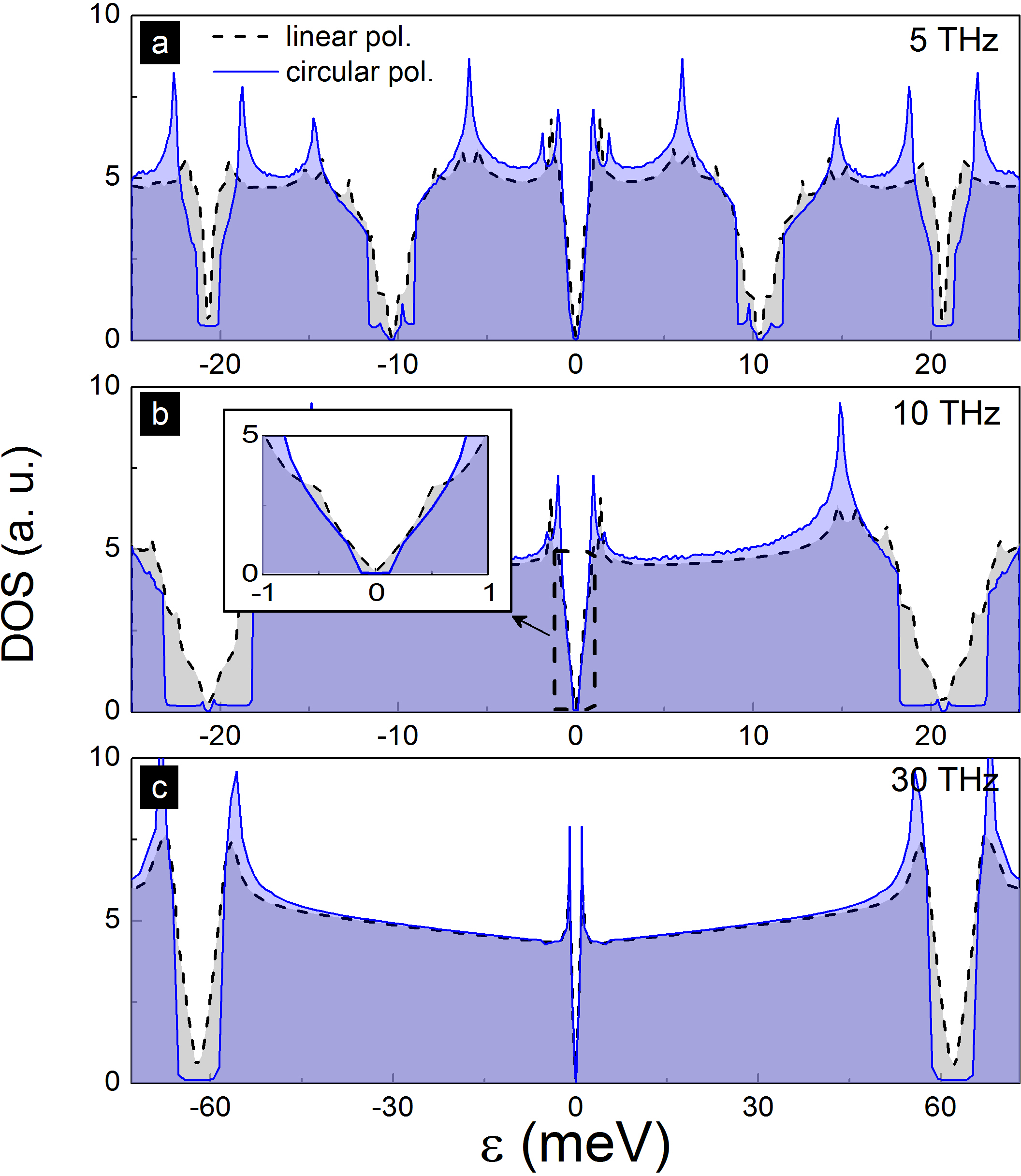}
\caption{DOS as a function of energy for (a) 5THz, (b) 10THz and (c) mid-infrared 30THz radiation. (a) and (b) correspond to a laser intensity of $0.5$ mW/$\mu$ $m^2$, (c) is computed for $10$ mW/$\mu$ $m^2$. Solid (dashed) lines are for circular (linear) polarization. Notice the change in the horizontal scale in (c). The structure induced by the TW becomes smoothened as the frequency increases. The inset in (b) is a zoom around zero energy.}
\label{fig-3}
\end{figure}

As one moves to higher frequencies, trigonal warping effects become less noticeable, though the gaps may become larger and therefore easier to observe experimentally. Figure \ref{fig-3} highlights this for three different frequencies a) $5$ THz, b) $10$ THz and c) $30$ THz (which corresponds to the mid-infrared range) for linearly (dashed line) and circularly (solid line) polarized light.

\subsection{Effective low-energy Hamiltonian description and topological considerations} 
Though more academic in nature, we now turn to an instructive analysis of the low-energy and topological properties of irradiated BLG. Our main fundamental question is: are there non-trivial laser-induced topological states to be expected in bilayer graphene? To such end, we are interested in obtaining an effective Hamiltonian to describe low energies electronic properties for low values of the light intensity in the spirit of Kitagawa and coworkers \cite{Kitagawa2011}. We will consider only the process when one photon is absorbed (emitted) and then re-emitted (re-absorbed), in this case applying the continued fraction method and retaining only the terms of order O($F^2$), where $F=eA/\hbar$ and in the following $\hbar=1$, the effective time-independent Hamiltonian can be expressed as:

\begin{equation}
H_{eff} = H_{0}+ V_{-1} \hat{G}(-1,\Omega) V_{+1} + V_{+1} \hat{G}(+1,\Omega) V_{-1}
\label{eq_Eff_Hamiltonian} 
\end{equation}
where $V_{\pm 1}=H^{(n,m)}$ for $n-m=\pm 1$ and $\hat{G}(n,\Omega)=\frac{1}{\epsilon + n \Omega -H_{0}}$ represents the propagator of a particle with $n$ photons.  For circularly polarized light, this results in the following effective Hamiltonian:  

\begin{equation}
H=\xi \left( \begin{array}{cccc}   
\frac{F^2 v_{3}^{2}}{\Omega}+\frac{F^2 v^{2} \Omega}{\gamma_{1}^{2}} & v_{3} \pi & 0 & v \pi^{\dagger}\\
v_{3} \pi^{\dagger} & -\frac{F^2 v_{3}^{2}}{\Omega}-\frac{F^2 v^{2} \Omega}{\gamma_{1}^{2}} & v \pi & 0\\
0 & v \pi^{\dagger} & -\frac{F^2 v^{2}}{\Omega} & \xi \gamma_{1}\\
v \pi & 0 & \xi \gamma_{1} & \frac{F^2 v^{2}}{\Omega}
\end{array} \right),
\label{eq_Ham4x4Eff} 
\end{equation}
where we have assumed $\gamma_{1} \gg \Omega \gg \epsilon$. All the terms in the diagonal should be multiplied by a factor $\eta=\pm 1$ to take into account left or right polarization of the light. Strikingly, this effective Hamiltonian resembles the Hamiltonian of bilayer graphene with a bias, but there are some subtle differences that a careful analysis reveals. One may argue that laser illumination introduces three ingredients: First, it breaks the intra-layer symmetry by introducing a term similar to Kane-Mele spin-orbit term ($F^2 v^2/ \Omega$)\cite{Kane_2005,Kitagawa2011} (if the layers were decoupled the system would have a gap solely due to this term); second, it breaks the inversion symmetry between the two layers (similar to a potential difference between layers), an effect which also opens a gap. And third,  when a graphene-based system with a gap is exposed to circularly polarized light an asymmetry between the valleys is expected due to the breaking of inversion symmetry, an effect similar to optical circular dichroism for valleys instead of spins.\cite{Wang2008,Inoue2011} The valley degree of freedom can be exploited generating valley dependent currents as we argue below. \\

The gap at $k=0$  is given by $2 \times( \frac{F^2 v_{3}^{2}}{\Omega}+\frac{F^2 v^{2} \Omega}{\gamma_{1}^{2}})$, the relative importance of these two terms is set by the frequency $\Omega$: for $\Omega$ in the THz range the trigonal warping term has a leading impact on the gap as previously noticed in the discussion of Fig. \ref{fig-2}.

To evaluate the topological properties of this effective Hamiltonian we reduce the previous $4\times4$ to a $2\times2$ Hamiltonian which describes the effective interaction between the non-dimer sites A1-B2. Considering as before $\gamma_{1} \gg \epsilon$ the new effective low energy Hamiltonian is given by:

\begin{equation}   
H= \varepsilon_{0} \left( \begin{array}{cc}   
\Delta & k_{-\xi}^{2}-  s \xi  k_{\xi} \\
k_{\xi}^{2}- s \xi k_{-\xi} & -\Delta \\
\end{array} \right),
\label{eq_Ham4x4Efflow} 
\end{equation}
where $\varepsilon_{0}=(\gamma_{3}/\gamma_{0})^{2}\gamma_{1}\approx 4 $ meV, $ \Delta= \eta \, F^{'2} (\frac{1}{\Omega^{'}}+\frac{\gamma_{3}^2 \Omega^{'}}{\gamma_{0}^2})$, $\xi=1(-1)$ for valley $K(K')$ , $k_{\pm} = (k_{x} \pm ik_{y})/k_{0}$, $k_{0}=2 \gamma_{3} \gamma_{1}/(\sqrt{3}a \gamma_{0}^{2})$, $a=0.246$ nm and $F^{'}$, $\Omega^{'}$ are now dimensionless parameters given in $k_{0}$ and $\varepsilon_{0}$ units respectively. The parameter $s$ takes values $(1,0)$ to include or not the trigonal warping. For values of $\Omega \ll \gamma_{1}$ the second term in the expression of $\Delta$ can be neglected and it gives a quite simple dependence of the gap with $F^{'}$ and $\Omega^{'}$, Gap=$2 \times \varepsilon_{0} \frac{F^{'2}}{\Omega^{'}}$. This expression shows an excellent agreement with numerical calculations in the frequency range considered to obtain Eq. \ref{eq_Ham4x4Eff}. 

From this effective Hamiltonian it is straightforward to calculate the Berry curvature and the Chern number \cite{Niu2010}. The curvature is given in polar coordinates by:   

\begin{equation}   
\Omega(k,\theta)=\frac{\xi \eta \Delta (4 k^2-s^2)}{2(\Delta^{2}+k^4+k^2s^2-2k^3 \xi s \ cos \, 3\theta)^{3/2}},
\label{eq_BerryCurvat} 
\end{equation}
and the integration gives an integer non-zero Chern number per valley, a quantum valley-Hall state \cite{Nota1}.  The Chern number has opposites values for the two valleys for a given handedness of polarization. A valley current will be proportional to the Berry curvature\cite{Niu1995}. Therefore a change in the handedness implies a change in the direction of the valley currents, as the sign of the Berry curvature changes. It provides an effective way to control these valley currents. There have been some proposals about this subject see for instance Ref.\cite{Schomerus2010,Martin2008}. On the other hand the structure of the Berry curvature reveals the impact of the trigonal warping: For low values of $\Delta$ the shape of the curvature shows a central dip with a topological charge Q=-1 and three peaks away from the center and separated $120^{o}$ with Q=1 each; in the $K'$ valley we have the opposite behavior. This segregation might have an impact on the edge currents of a system based on bilayer graphene and energies in the Teraherz range \cite{Morell2011}.

A completely different picture is obtained from irradiating bilayer graphene with linearly polarized light, following the same procedure as before, one obtains a gap at $k=0$, with a peculiar behavior, it does not depend explicitly on $\Omega$ neither on $\gamma_{3}$, Gap=$2 \times \frac{F^2 v^2}{\gamma_{1}}$. The Chern number equals zero in every valley, thus the states are topologically trivial.

\section{Conclusions}

In summary, the effects of a laser with frequencies ranging from THz to the mid-infrared on the electronic structure of bilayer graphene are analyzed, highlighting the appeareance of laser-induced gaps and their dependence with the light polarization as well as the strong influence of trigonal warping corrections. For radiation in the THz range, trigonal warping in bilayer graphene tends to decrease the size of the dynamical gaps at  $ \pm\hbar \Omega /2$, this is very different from the case of monolayer graphene where trigonal warping effects are much weaker \cite{Calvo2011}. Furthermore, we obtain a time-independent effective Hamiltonian which serves as a starting point for the determination of the topological properties of the associated low-energy states. We find that while for both polarizations there is a small gap at zero energy, their topological origin is different: The Chern number in the presence of linearly polarized light equals zero, a trivial insulator, while it is a non zero integer, a quantum valley-Hall insulator, when the light is circularly polarized. Though more difficult to observe experimentally than the dynamical gaps, further work in this direction may open promising prospects for exploiting the valley degree of freedom in graphene-based structures.

\textit{Acknowledgments.} ESM acknowledges support from DGIP, UTFSM. LEFFT acknowledges funding by SeCyT-UNC, ANPCyT-FonCyT, and the support from the Alexander von Humboldt Foundation and the ICTP of Trieste. We acknowledge discussions with D. Soriano Hernandez and H. L. Calvo.


\begin{thebibliography}{99}
\bibitem{note} The photothermal response of the system is not considered here. In an actual experiment further measurements of their different polarization and wavelength dependences may help to distinguish them.
\end{thebibliography}


\end{document}